\documentclass[journal]{IEEEtran}
\usepackage{amsfonts,amsmath,amssymb, amsthm} 
\usepackage{siunitx}
\usepackage{microtype}
\usepackage{algorithmic}
\usepackage{algorithm}
\usepackage{array}
\usepackage[caption=false,font=normalsize,labelfont=sf,textfont=sf]{subfig}
\usepackage{textcomp}
\usepackage{stfloats}
\usepackage{url}
\usepackage{verbatim}
\usepackage{graphicx}
\usepackage{cite}
\usepackage{booktabs}
\usepackage[hidelinks]{hyperref}
\usepackage{url}
\usepackage{multirow}
\usepackage{xspace}
\newcommand{\resolved}[1]{}

\newcommand{\ours}{AudioLM\xspace}
\newcommand{\wvbert}{w2v-BERT\xspace}
\newcommand{\waveform}{x}
\newcommand{\waveformhat}{\hat{x}}
\newcommand{\tokens}{y}
\newcommand{\semantictokens}{z}

\newcommand{\ssmatrix}{Y}

\newcommand{\semantictokenshat}{\hat{z}}

\newcommand{\numsamples}{T}
\newcommand{\numtokens}{T'}
\newcommand{\numtokensacoustic}{{T_A}}
\newcommand{\numtokenssemantic}{{T_S}}
\newcommand{\numquantizers}{Q}
\newcommand{\numtopquantizers}{Q'}

\newcommand{\codebooksize}{N}
\newcommand{\codebooksizesemantic}{K}

\newcommand{\enc}{\text{enc}}
\newcommand{\dec}{\text{dec}}

\usepackage{pgfplotstable}

\pgfplotstableset{col sep=comma}
\pgfplotstableset{header=true}
\pgfplotstableset{
every head row/.style={
before row=\toprule,
after row=\midrule,
},
every last row/.style={
after row=\bottomrule,
},}

\usepackage[english, status=draft]{fixme}
\fxusetheme{colorsig}
\fxsetup{theme=color, mode=multiuser}
\FXRegisterAuthor{zb}{azb}{Zal\'an} % Now you can do \nzbnote, \zbwarning
\FXRegisterAuthor{rm}{arm}{Rapha\"el}
\FXRegisterAuthor{dv}{adv}{Damien}
\FXRegisterAuthor{ek}{aek}{Eugene}
\FXRegisterAuthor{mt}{amt}{Marco}
\FXRegisterAuthor{nz}{anz}{Neil}

\begin{document}

\title{\ours{}: a Language Modeling Approach to Audio Generation}

\author{Zal\'an Borsos, Rapha\"el Marinier, Damien Vincent, Eugene Kharitonov, Olivier Pietquin,\\Matt Sharifi, Dominik Roblek, Olivier Teboul, David Grangier, Marco Tagliasacchi, Neil Zeghidour \thanks{Google Research}}

\maketitle

\begin{abstract}
\looseness=-1
We introduce~\ours{}, a framework for high-quality audio generation with long-term consistency. \ours{} maps the input audio to a sequence of discrete tokens and casts audio generation as a language modeling task in this representation space. We show how existing audio tokenizers provide different trade-offs between reconstruction quality and long-term structure, and we propose a hybrid tokenization scheme to achieve both objectives. Namely, we leverage the discretized activations of a masked language model pre-trained on audio to capture long-term structure and the discrete codes produced by a neural audio codec to achieve high-quality synthesis. By training on large corpora of raw audio waveforms, \ours{} learns to generate natural and coherent continuations given short prompts. When trained on speech, and without any transcript or annotation, \ours{} generates syntactically and semantically plausible speech continuations while also maintaining speaker identity and prosody for unseen speakers. Furthermore, we demonstrate how our approach extends beyond speech by generating coherent piano music continuations, despite being trained without any symbolic representation of music.
\end{abstract}

% \begin{IEEEkeywords}
% Article submission, IEEE, IEEEtran, journal, \LaTeX, paper, template, typesetting.
% \end{IEEEkeywords}

\section{Introduction}
\looseness=-1

\looseness=-1
\IEEEPARstart{A}{udio} signals, be they speech, music or environmental sounds, involve multiple scales of abstractions. For instance, speech can be analyzed at a very local acoustic or phonetic level but also in terms of prosody, syntax, grammar, or semantics. Music also follows a long-term structure, while being composed of highly non-stationary acoustic signals. When it comes to audio synthesis, these multiple scales interact in such a way that achieving high audio quality while displaying high-level consistency remains a challenge, in particular in the absence of strong supervision. 

\looseness=-1
Recent audio synthesis models have achieved nearly veridical signal quality by leveraging methods such as autoregressive waveform modeling~\cite{oord2016wavenet, kalchbrenner2018wavernn}, adversarial training~\cite{kumar2019melgan, kong2020hifigan, tagliasacchi2020seanet} or diffusion~\cite{diffwave, wavegrad}. Yet, when not provided with strong conditioning (e.g., linguistic features, a MIDI sequence), even powerful models like WaveNet~\cite{oord2016wavenet} generate unstructured audio, such as babbling speech.
Language models, on the other hand, have demonstrated their ability to model high-level, long-term structure for different content types, and the consequent advances in text~\cite{gpt3, gopher, palm} and image generation~\cite{vit-vqgan2021, parti} have paved the way towards synthesis of natural audio that remains intelligible and consistent over time. An important step in that direction, coined as ``textless NLP'', has been recently achieved for unconditioned speech generation~\cite{zerospeech21, lakhotia2021generative}. In particular, Lakhotia et al.~\cite{lakhotia2021generative} show that a Transformer~\cite{attentionvaswani} trained on discretized speech units can generate coherent speech without relying on textual annotations. Yet, the acoustic diversity and the quality remain limited: the model is trained on clean speech only and synthesis is restricted to a single speaker.

\looseness=-1
In this work, we introduce \ours{}, a framework that enables high-quality audio generation with long-term coherent structure, as demonstrated by our experiments on both speech and piano music continuation. We achieve this objective by combining recent advances in adversarial neural audio compression~\cite{soundstream}, self-supervised representation learning~\cite{w2vbert} and language modeling~\cite{roberts2022t5x}. Specifically, starting from raw audio waveforms, we first construct coarse \emph{semantic tokens} from a model pre-trained with a self-supervised masked language modeling objective~\cite{bert}. 
Autoregressive modeling of these tokens captures both local dependencies (e.g., phonetics in speech, local melody in piano music) and global long-term structure (e.g., language syntax and semantic content in speech; harmony and rhythm in piano music). However, these tokens lead to poor reconstruction.
To overcome this limitation, in addition to semantic tokens, we rely on fine-level \emph{acoustic tokens} produced by a SoundStream neural codec~\cite{soundstream}, which capture the details of the audio waveform and allow for high-quality synthesis. Training a language model to generate both semantic and acoustic tokens leads simultaneously to high audio quality and long-term consistency. In summary, we make the following contributions:
\begin{itemize}
\setlength\itemsep{0.3em}
    \item We propose \ours{}, a framework for audio generation that combines semantic and acoustic tokens in a hierarchical fashion to achieve long-term consistency and high quality.
    \item We compare the semantic tokens extracted from a pre-trained \wvbert~\cite{w2vbert} and the acoustic tokens from SoundStream~\cite{soundstream} on a speech dataset, and we show that they complement each other in terms of phonetic discriminability and reconstruction quality.
    \item We demonstrate the ability of \ours{} to generate coherent speech in terms of phonetics, syntax and semantics, without relying on textual annotations. Moreover, when conditioned on a prefix (or \textit{prompt}) of only \SI{3}~seconds of speech from a speaker not seen during training, \ours{} produces consistent continuations while maintaining the original speaker voice, prosody and recording conditions (e.g., level of reverberation, background noise).
    \item We show that \ours{} is also suited for music generation. When training on piano recordings, it generates convincing continuations that are coherent with the prompt in terms of melody, harmony, tone and rhythm.
    \item We acknowledge the potential risks associated with the use of generative models that enable speech continuation, and we mitigate these risks by training a classifier that can detect synthetic speech generated by \ours{} with very high accuracy.
\end{itemize}

We encourage the reader to listen to the samples produced by \ours{} in the accompanying material.\footnote{\label{webpage}\url{https://google-research.github.io/seanet/audiolm/examples}}

\section{Related work}
\looseness=-1
\noindent
\textbf{High-fidelity neural audio synthesis.} Recent years have seen tremendous progress in the quality of audio generated by neural networks, largely attributed to the introduction of objective functions that improve over simple waveform regression. In particular, WaveNet~\cite{oord2016wavenet} introduced an autoregressive classification approach to speech synthesis, with quality that significantly outperformed traditional concatenative and parametric approaches at the cost of slow inference. While WaveNet inspired more computationally efficient alternatives such as WaveRNN~\cite{kalchbrenner2018wavernn} or parallel WaveNet~\cite{parallel_wavenet}, a significant paradigm shift occurred with the introduction of adversarial audio generation~\cite{kumar2019melgan, gantts, kong2020hifigan}, which enables high fidelity generation without any autoregressive component. Moreover, combining such high-quality synthesis systems with differentiable quantization~\cite{vqvae, vqvae2, softhardquantization}, allows training end-to-end neural codecs~\cite{kankanahalli2018speechdnn, zhen2019cascaded, soundstream, petermann2021harp} by compressing activations in a bottleneck layer. \ours{} leverages the tokens produced by a SoundStream neural codec~\cite{soundstream}, not as intermediate representations for lossy reconstruction, but rather as targets for a sequence modeling task operating at a lower sampling rate, which can be decoded back to audio at the original sampling rate.

\looseness=-1
\noindent
\textbf{Self-supervised learning of audio representations.} While neural audio synthesis typically focuses on modeling fine details of the signal, most self-supervised learning approaches rather aim at discovering high-level representations that correlate with coarse, symbolic features (e.g., phonemes, musical notes, class labels). This is typically achieved by proposing proxy objectives that do not rely on any transcript or label, but rather exploit regularities in the structure of the audio signals. Among these approaches, contrastive training learns representations for which pairs of positive examples are closer to each other than negative pairs. Positive pairs can be, for example, two segments that are close temporally~\cite{cpc,wav2vec2,cola} or two augmented views of the same sequence~\cite{contrastive_kharitonov}. 

Another line of work, inspired by NLP systems pre-training~\cite{bert,roberts2022t5x}, has explored the  discretization of audio signals into a finite vocabulary of tokens to serve as targets for masked language modeling pre-training~\cite{bert}, i.e. predicting long contiguous spans of masked tokens from a wide context. The discretization strategy is critical to the downstream performance of such models. Popular quantization strategies include quantizing representations optimized for future time step prediction~\cite{vqwav2vec}, starting from quantizing low-level audio features followed by iterations of quantization target refinement~\cite{hubert}, and jointly learning the quantization along with the masked language model~\cite{w2vbert}. The discriminative nature of these contrastive and predictive objectives, as well as the fact that they require exploiting long-term dependencies, allow learning representations that encode coarse, high-level information about the signal (e.g., phonemes and word identity when trained on speech~\cite{pasad2021}). These representations are thus particularly useful for discriminative downstream tasks such as speech recognition~\cite{hubert} or audio classification~\cite{cola}. However, as they are not optimized to encode fine details of original audio signals, they are poorly invertible and thus not directly usable for synthesis. \ours{} avoids this limitation by leveraging these high-level representations as a conditioning signal that carries semantic information and guides the prediction of high-quality acoustic tokens.

\looseness=-1
\noindent
\textbf{Generating natural signals with language models.}  Neural language models have demonstrated remarkable abilities for tasks as diverse as open-ended dialog modeling~\cite{lamda}, code completion~\cite{openaicodex} or even solving integrals and differential equations~\cite{symbolic_math_lample}. The key underlying mechanism of the best of these models is self-attention~\cite{attentionvaswani}, which is suitable for modeling rich and complex long-range dependencies but, in the standard form, has a computational cost that grows quadratically with the length of the input sequences. This cost is acceptable for sequences of up to $10^3$ tokens~\cite{t5}, however, it prevents modeling natural signals in their raw form (for example, modeling a $512\times512$ image at the pixel level). While several works have explored efficient alternatives to self-attention~\cite{routing_transformer,performer,perceiverar2022}, another solution to this scaling problem is to work with mappings of the natural signals to a compact, discrete representation space. A common approach is to model the representations in this space with an autoregressive Transformer, whose predictions are then mapped back to the original signal space. This approach has been used to generate high-resolution images~\cite{tamingtransformers,vit-vqgan2021,parti} and long videos~\cite{tats}. 

For audio, Jukebox~\cite{jukebox} adopts a hierarchical approach to generate tokens at various temporal resolutions which are then combined to reconstruct music. Another notable line of work is ``textless NLP''~\cite{zerospeech21,lakhotia2021generative,pgslm,textlessLib}, which models language directly in the speech domain, without any transcription, by training autoregressive generative models of low-bitrate audio tokens~\cite{cpc,hubert}. While Jukebox and GSLM~\cite{lakhotia2021generative} show high temporal coherence (e.g., spoken language generated by GSLM is meaningful), their audio quality remains limited: the music generated by Jukebox displays significant artifacts, while the speech sampled from GSLM is limited to a single speaker in a clean setting. This is unlike Perceiver AR~\cite{perceiverar2022}, which trains an autoregressive model on the discrete codes of a high-bitrate SoundStream~\cite{soundstream} codec. The model can then generate piano music of high signal-level quality; however the temporal structure of the generated sequences can be further improved. \ours{} tackles both challenges of long-term coherence and high-quality by combining semantic and acoustic tokens in a generative framework. This leads to improvements over GSLM by generating speech continuations that preserve the original speaker's identity and intonation, as well as extending audio continuation beyond speech by generating piano sequences with high-level coherence.

\section{Model}

\begin{figure}[t]
\centering
\includegraphics[width=0.48\textwidth]{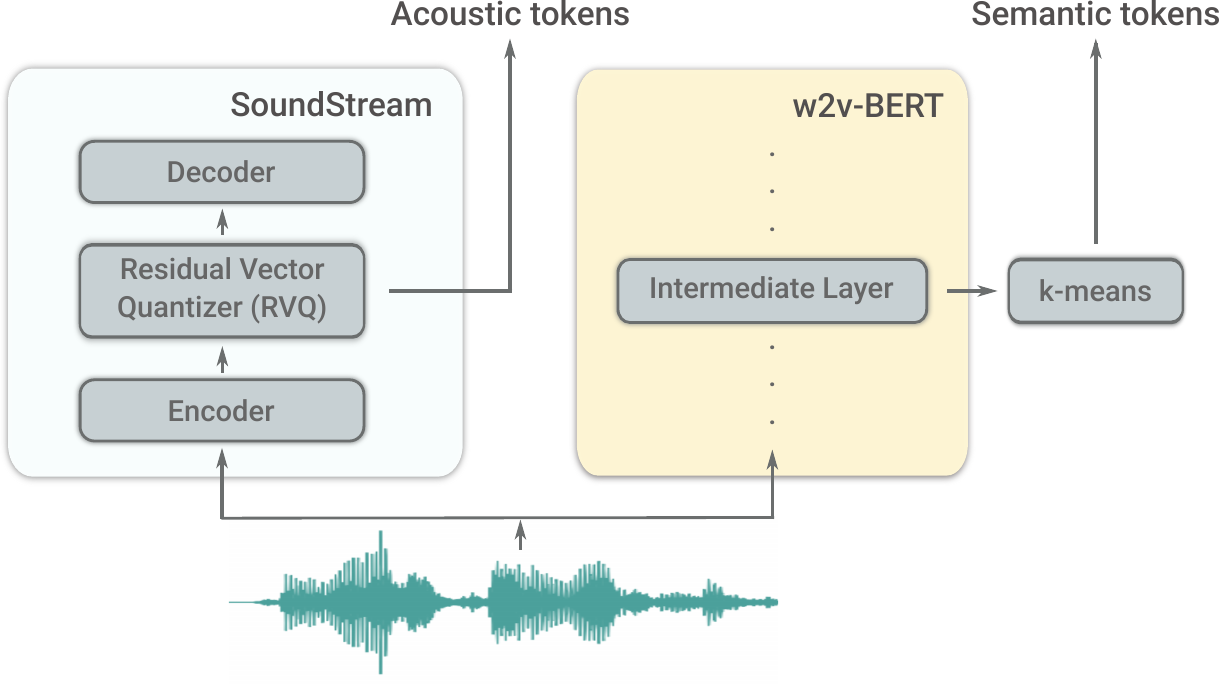}
\caption{Overview of the tokenizers used in \ours{}. The acoustic tokens are produced by SoundStream~\cite{soundstream} and enable high-quality audio synthesis. The semantic tokens are derived from representations produced by an intermediate layer of  w2v-BERT~\cite{w2vbert} and enable long-term structural coherence.
\label{fig:audiolm-tokenizers} } 
\end{figure}

In this section, we first describe the components of our framework, together with the representation and modeling challenges in audio generation. We address these challenges by proposing a hybrid tokenization scheme together with a multi-stage Transformer-based language model operating on the proposed tokens. 

\subsection{Components}
We consider a single channel audio sequence $\waveform \in \mathbb{R}^{\numsamples}$, which is processed by the following three components of the \ours{} framework:
\begin{itemize}
    \item A tokenizer model, which maps $\waveform$ into a sequence $h =\enc(x)$, $h=(h_1,\dots,h_{T'})$ of discrete tokens from a finite vocabulary, with $T'\ll T$.
    
    \item A decoder-only Transformer language model that operates on the discrete tokens $\tokens$, trained to maximize the likelihood $\prod_{t=1}^{T'}p(h_{t} | h_{<t})$. 
    At inference time, the model predicts the token sequence $\hat{h}$ autoregressively. 
    
    \item A detokenizer model, which maps the sequence of predicted tokens back to audio, producing the waveform $\waveformhat = \dec(\hat{h})$.
\end{itemize}

It is important to emphasize the following aspects: i) the number of tokens $\numtokens$ is typically 2-3 orders of magnitude smaller than $\numsamples$. This is critical to significantly increase the temporal context size of the language model, since the computational complexity of standard self-attention grows quadratically with respect to the sequence length; ii) the tokenizer and detokenizer are pre-trained and frozen ahead of training the language model, which decouples the tokenizers and the language model and simplifies the training setup.

\subsection{Trade-offs of discrete audio representations} \label{sec:tradeoffs}
The tokenizer and detokenizer models allow us to operate on discrete audio representations. On the one hand, we want to be able to reconstruct audio waveforms at high quality, which introduces a lower bound on the bitrate and hence on the length of the token sequence. On the other hand, we aim at obtaining a compact representation that captures long-term dependencies. To reconcile these conflicting requirements, we rely on a combination of acoustic and semantic tokens, which are illustrated in Figure~\ref{fig:audiolm-tokenizers}. In this tokenization scheme, the semantic tokens enable long-term structural coherence, while modeling the acoustic tokens conditioned on the semantic tokens enables high-quality audio synthesis.

\looseness -1
We compute acoustic tokens using SoundStream~\cite{soundstream}, a state-of-the-art neural audio codec, which significantly outperforms non-neural codecs like Opus and EVS at low bitrates. SoundStream adopts a convolutional encoder to map the input waveform to a sequence of embeddings, whose sampling rate is significantly lower than the sampling rate of the original audio. We configure SoundStream to produce embeddings at 50 Hz (one every 20 ms) for input waveforms at 16 kHz. This is a 16000 / 50 = 320-fold reduction in the sampling rate. Each embedding is discretized using a residual vector quantizer (RVQ), which consists of a hierarchy of $\numquantizers$ vector quantizers, each using a vocabulary of $\codebooksize$ symbols. For example,  using $\codebooksize = 1024$, $\numquantizers=4$ results in a bitrate of 2000\,bps ($50 \cdot 4 \cdot \log_2 1024 $). Hence, the input audio samples $\waveform$ are represented by a matrix $\ssmatrix \in \{1,\dots,\codebooksize\}^{\numtokensacoustic\times\numquantizers}$ of codebook symbols, with $\numtokensacoustic=\numsamples / 320$. Then, the convolutional decoder of SoundStream maps this discrete representation to real-valued embeddings and then reconstructs the waveform. The codec achieves high quality by being trained end-to-end with a  combination of reconstruction and adversarial losses. 

We compute semantic tokens using \wvbert~\cite{w2vbert}, a recently proposed model for learning self-supervised audio representations. When trained on large speech corpora, \wvbert learns to map the input audio waveform to a rich set of linguistic features. This is achieved by training a 0.6B-parameter Conformer-based model~\cite{conformer} using a combination of two self-supervised objectives: a masked language modeling (MLM) loss and a contrastive loss. While this model can be fine-tuned for discriminative tasks such as speech recognition or speech-to-text translation~\cite{xtremes}, \ours{} rather leverages the representations of the pre-trained \wvbert to model long-term temporal structure in a generative framework. To this end, we select an intermediate layer of the MLM module of \wvbert and compute embeddings at this level. We train a $k$-means with $\codebooksizesemantic$ clusters on these embeddings and use the centroid indices as semantic tokens. We found that normalizing \wvbert embeddings such that each dimension has zero mean and unit variance before clustering significantly improves their phonetic discriminability. \wvbert performs downsampling along the temporal dimension, so that real-valued 1024-dimensional feature vectors are computed at a sampling rate of 25 Hz (one every 40 ms). Hence, the input audio samples $\waveform$ are transformed into a sequence of semantic tokens $\semantictokens=(z_1,\dots,z_{T_S}) \in \{1,\dots,\codebooksizesemantic\}^\numtokenssemantic$ with $\numtokenssemantic = \numsamples / 640$. For example, when $T=16000$,  $\codebooksizesemantic = 1024$, this results in a bitrate equal to 250\,bps. We note that our proposal for the extraction of semantic tokens from w2v-BERT resembles the token extraction from HuBERT in prior works~\cite{lakhotia2021generative, pgslm}.

\looseness -1
To motivate our hybrid tokenization scheme, we contrast the different properties of the acoustic tokens obtained from SoundStream, and the semantic tokens obtained from \wvbert, by comparing them in terms of audio quality reconstruction and phonetic discriminability. We evaluate the reconstruction quality by training a SoundStream decoder to reconstruct audio from tokens. We then compute the ViSQOL score~\cite{hines2015visqol, chinen2020visqol}, a computational proxy for perceived similarity between a reference audio and its reconstruction. In particular, we use the ``speech'' mode, which operates on 16 kHz signals.

\looseness=-1
We measure phonetic discriminability in terms of ABX error rate~\cite{abx}. It is a distance-based metric that considers a set of phoneme trigrams which only differ in the central phoneme (e.g., ``bit'' vs.\ ``bet''). ABX error rate measures how often a random instance X of a trigram (``bit'') is closer to an instance B of another trigram (``bet'') rather than to a different instance A of the same trigram (``bit''). We consider cases where all three sounds A, B, and X are uttered by the same speaker (within-speaker) and where A and B are uttered by the same speaker and X is coming from a different speaker (across-speaker)~\cite{schatz2016abx}. To allow uniform comparison across the two representations, we represent speech using residual vector-quantized embeddings, where each frame is represented by its corresponding centroid for \wvbert or by the output of a SoundStream quantizer. We calculate ABX  using scripts published with the Libri-Light dataset~\cite{librilight} with the default settings and report scores obtained on LibriSpeech dev-clean~\cite{librispeech}.
Table~\ref{tab:token_benchmark} shows that acoustic tokens provide a good reconstruction quality (ViSQOL of 3.3  for 2000\,bps, 3.9 for 6000\,bps), but poor phonetic discriminability. Conversely, semantic tokens extracted from the 7th layer from the MLM module of \wvbert  significantly improve phonetic discriminability, but they do not attain high reconstruction quality, even when matching the bitrate of the acoustic tokens. 

\looseness=-1
Consequently, achieving both high quality and long-term consistency with only one of the tokenizers is challenging. To illustrate this point further, we can model the sequences of one of the token types and inspect the properties of the resulting model. We perform this on the acoustic tokens, since the semantic tokens only allow for poor audio synthesis. We train a decoder-only Transformer on the sequence of acoustic tokens, by flattening $Y$ in a row-major order to a sequence of tokens $\tokens + o$ of length $\numtokensacoustic\cdot\numquantizers$, where $\tokens = (y_{1}^{1}, y_{1}^{2}, \ldots, y_{1}^{Q}, y_{2}^{1}, \ldots, y_{\numtokensacoustic}^{\numquantizers})$, $y_{t}^{q}$ is the token produced by the $q$-th quantizer for the $t$-th time step, and $o=(o_1,o_2, \dots,o_{\numtokensacoustic\cdot\numquantizers})$ is the vector of offsets for creating unique token indices for the $Q$ layers of the residual vector quantizer, with $o_i = (i - 1 \; \textup{mod} \; Q) \cdot N$. In the following, we omit the offsets from the notation and assume proper offsetting implicitly. Using the model trained only on the acoustic tokens, we sample speech continuations from a prompt of \SI{4}~seconds. While both the recording conditions and the speaker identity from the prompt are preserved, the linguistic content is inconsistent, and often akin to babbling (see ``Generation without semantic tokens'' in the accompanying material\textsuperscript{\ref{webpage}}).

\begin{table*}[t]
\centering
\caption{Comparison of token types in terms of phonetic discriminability within and across speakers (lower is better) and reconstruction quality (higher is better). Phonetic discriminability is measured by ABX, while reconstruction quality is reported in ViSQOL units.}
\label{tab:token_benchmark}
\begin{tabular}{lccc}
\toprule
Tokenization &  Bitrate & Phonetic discriminability within/across ($\downarrow$) & Reconstruction quality ($\uparrow$) \\
\midrule
\multirow{2}{*}{Semantic (\wvbert) } & \SI{250}{bps}  & $6.7$ / 7.6 & $1.1$\\
& \SI{6000}{bps} & 5.6 /  6.2& $1.4$ \\
\midrule
\multirow{2}{*}{Acoustic (SoundStream) } & \SI{2000}{bps}  & $22.4$ / $28.7$ & $3.3$\\
& \SI{6000}{bps} & $17.8$ / $26.6$ & $3.9$ \\
\bottomrule
\end{tabular}
\vspace{-0.2cm}
\end{table*}

\begin{figure*}[t]
\centering
\includegraphics[width=\textwidth]{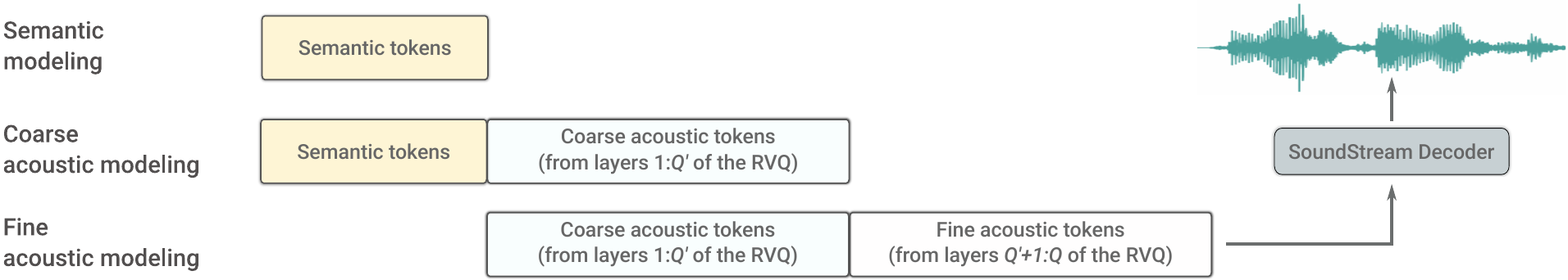}
\caption{The three stages of the hierarchical modeling of semantic and acoustic tokens in \ours{}: i) semantic modeling for long-term structural coherence, ii) coarse acoustic modeling conditioned on the semantic tokens and iii) fine acoustic modeling. With the default configuration, for every semantic token there are $2 \numtopquantizers$ acoustic tokens in the second stage and $2 (\numquantizers - \numtopquantizers)$ tokens in the third stage. The factor of 2 comes from the fact that the sampling rate of SoundStream embeddings is twice as that of the \wvbert embeddings.
\label{fig:audiolm-stages} } 
\end{figure*}

\subsection{Hierarchical modeling of semantic and acoustic tokens}
The observations in the previous section suggest that, by modeling both semantic and acoustic tokens within the same framework, the semantic tokens would ensure long-term consistency (by capturing linguistic content for speech, melody and rhythm for music), while the acoustic tokens would ensure high-quality audio synthesis (by capturing the acoustic details). We build the \ours{} framework on this hypothesis. Concretely, we adopt a hierarchical approach, by first modeling the semantic tokens for the entire sequence, and then use these as conditioning to predict the acoustic tokens. This approach has two main advantages: i) the hierarchical modeling reflects the conditional independence assumption that semantic tokens are expected to be conditionally independent from past acoustic tokens given past semantic tokens, that is, $p(z_t | z_{<t}, y_{<t}) \approx p(z_t | z_{<t})$; ii) the token sequence per stage is reduced compared to alternatives such as modeling the interleaved sequence of semantic and acoustic tokens, allowing for computationally more efficient training and inference.

\ours{} performs three subsequent stages, as illustrated in Figure~\ref{fig:audiolm-stages}. In all stages, we use a separate decoder-only Transformer trained for predicting next tokens given all previous ground-truth tokens in the corresponding stage.

\noindent
\textbf{Semantic modeling.} The first stage models $p(\semantictokens_t | \semantictokens_{<t})$, the autoregressive prediction of semantic tokens to capture long-term temporal structure.

\noindent
\textbf{Coarse acoustic modeling.} The second stage proceeds analogously on the acoustic tokens, but it only predicts the acoustic tokens from the coarse $\numtopquantizers$ SoundStream quantizers, conditioned on the semantic tokens. Due to residual quantization in SoundStream, the acoustic tokens have a hierarchical structure: tokens from the coarse quantizers recover acoustic properties like speaker identity and recording conditions, while leaving only the fine acoustic details to the fine quantizer tokens, which are modeled by the next stage. We rely on the simple approach of flattening the acoustic tokens in a row-major order to handle their hierarchical structure. Consequently, the second stage models $p(\tokens_t^q | \semantictokens, \tokens_{<t}^{\leq \numtopquantizers}, \tokens_{t}^{< q})$, for $q \leq \numtopquantizers$, where the corresponding token sequence is $(z_1,z_2, \dots, z_{T_S}, y_1^1,y_1^2,\dots,y_1^{\numtopquantizers}, y_2^1,y_2^2,\dots,,y_2^{\numtopquantizers},\dots,y_{T_A}^{\numtopquantizers})$, with $y_1^1$ being the first token predicted during training.

\noindent
\textbf{Fine acoustic modeling.} The third stage operates on acoustic tokens corresponding to the fine quantizers, using the $\numquantizers'$ coarse tokens as conditioning and modeling the conditional probability distribution  $p(\tokens_t^q | \tokens^{\leq \numtopquantizers}, \tokens_{<t}^{> \numtopquantizers}, \tokens_{t}^{<q})$ for $q > \numtopquantizers$. That is, $\tokens_t^q$ is predicted based on all tokens corresponding to the coarse $\numquantizers'$ quantizers, followed by the fine $\numquantizers-\numtopquantizers$ quantizers at previous time steps, together with the already decoded tokens at the current time step corresponding to the coarser quantizers. In this stage, we further improve audio quality, removing the lossy compression artifacts that remain after the second stage. 

Although the second and third stage could be merged into a single stage, we adopt the solution with two separate stages to limit the sequence length that the model has to process at once. First, considering that fine acoustic tokens are conditionally independent from semantic tokens when conditioned on coarse acoustic tokens, the third stage can ignore the semantic tokens, which reduces the total sequence length. Moreover, under the assumption that the fine acoustic details are determined locally by the coarse acoustic tokens, we perform the third stage on batches of non-overlapping audio chunks of 3 seconds, allowing us to scale this stage independently of the target audio sequence length as well as to use more residual quantization layers $Q$ to achieve higher quality.

\subsection{Inference} \label{sec:model-inference}
After training, we can generate audio with \ours{} as detailed below. Depending on the conditioning signal used, we obtain different forms of generation. 

\noindent
\textbf{Unconditional generation.} In this setting, we sample unconditionally all semantic tokens $\semantictokenshat$, which we then use as conditioning for acoustic modeling. The samples in the accompanying material\textsuperscript{\ref{webpage}} show that the model generates diverse, syntactically and semantically consistent linguistic content, with varying speaker identity, prosody, acoustic conditions. Section~\ref{sec:linguistic-content} furthermore validates quantitatively the lexical and syntactic knowledge of the model.

\looseness -1
\noindent
\textbf{Acoustic generation.} In this setting, we use the ground-truth semantic tokens $\semantictokens$ extracted from a test sequence $\waveform$ as conditioning to generate the acoustic tokens. Sections \ref{sec:exp-semantic} and \ref{sec:exp-acoustic} show that, in this case, the generated audio sequences still vary in speaker identity but the content of the spoken sentence remains the same, matching the ground-truth transcript of $\waveform$. This shows that the semantic tokens capture the semantic content.

\looseness -1
\noindent
\textbf{Generating continuations.} Our main application of interest is generating continuations from a short prompt $\waveform$. To do so, we first map the prompt to the corresponding semantic tokens $\semantictokens_{\leq t_s}$ and to the coarse acoustic tokens $\tokens_{\leq t_a}^{\leq \numtopquantizers}$. The first stage generates $\semantictokenshat_{>t_s}$, the continuation of semantic tokens autoregressively based on the conditioning $\semantictokens_{\leq t_s}$. In the second stage, we concatenate the entire semantic token sequence $(\semantictokens_{\leq t_s}, \semantictokenshat_{>t_s})$ along with the coarse acoustic tokens of the prompt $\tokens_{\leq t_a}^{\leq \numtopquantizers}$ and feed it as conditioning to the coarse acoustic model, which then samples the continuations of the corresponding acoustic tokens. In the third stage, we process the coarse acoustic tokens with the fine acoustic model. Finally, we feed both the prompt and the sampled acoustic tokens to the SoundStream decoder to reconstruct a waveform $\waveformhat$. Section~\ref{sec:prompt-continuation} shows that, when prompted with only 3 seconds of speech from an unseen speaker, \ours{} generates continuations that are hardly distinguishable from the original voice. Moreover, Section~\ref{sec:piano-continuation} demonstrates the performance of AudioLM beyond speech, by continuing piano performances.

\section{Experiments}
In order to showcase the general applicability of the \ours{} framework, we consider two tasks from different audio domains: 
\begin{itemize}
    \item \textit{Speech continuation}, where the model is expected to keep the speaker identity, prosody and recording conditions of the prompt and produce new content, which is syntactically correct and semantically consistent.
    \item \textit{Piano continuation}, where the model is expected to generate piano music, which is coherent with the prompt in terms of melody, harmony and rhythm.
\end{itemize}

\looseness -1
As the speech and piano prompts we use for evaluation are respectively from unseen speakers and unseen performances, generating consistent continuations requires \ours{} to generalize beyond training data. We furthermore conduct experiments that provide empirical support for our hierarchical approach and shed light on the properties of the semantic and acoustic tokens. In order to mitigate the potential misuse of our framework, we provide an effective method for detecting speech generated by \ours{}.

\subsection{Datasets} \label{sec:datasets}
For speech, all components of \ours{} (SoundStream, \wvbert, the $k$-means quantizer for \wvbert embeddings, the decoder-only Transformers) are trained on the unlab-60k train split of Libri-Light~\cite{librilight}, consisting of 60k hours of English speech. While previous works~\cite{lakhotia2021generative, pgslm} use the 6k-hour clean subset~\cite{librilight-6k} of Libri-Light for training the language model, \ours{} shows strong performance when trained on the more diverse and noisy unlab-60k subset. The increased robustness to the quality of the training data reduces the data preparation effort needed to apply our framework.

\begin{figure}[t]
\centering
\subfloat{
\includegraphics[width=0.225\textwidth]{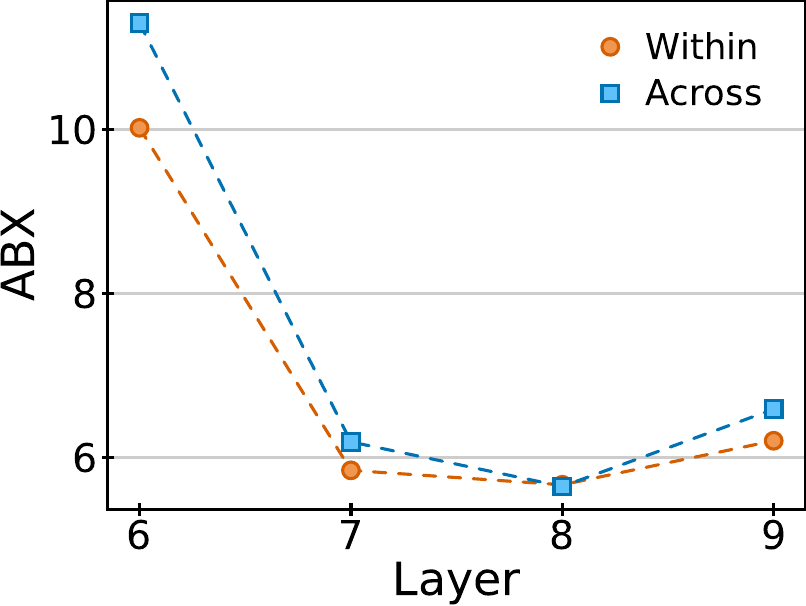}
\label{fig:abx}}
\subfloat{
\includegraphics[width=0.23\textwidth]{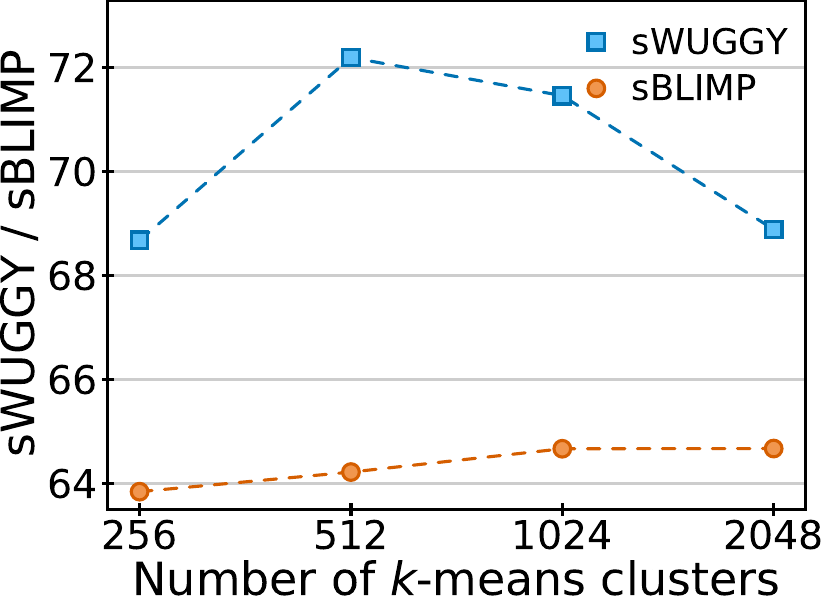}
\label{fig:swuggy-sblimp}}

\caption{Left: ABX ($\downarrow$) scores achieved by the (unquantized) embeddings extracted from different layers of the MLM module of \wvbert. Right: Scores on the development sets of sWUGGY ($\uparrow$) and sBLIMP ($\uparrow$) obtained with different numbers of $k$-means cluster centers for layer 7.}
\label{fig:abx-swuggy-sblimp}
\end{figure}

\subsection{Model selection, training and inference}
\looseness -1
\textbf{Semantic tokens.} We use \wvbert XL~\cite{w2vbert}  (0.6B parameters) and adopt a set of heuristics for choosing the intermediate layer to quantize and the number of $k$-means clusters $\codebooksizesemantic$. Namely, we inspect ABX, sWUGGY and sBLIMP scores (see Sections \ref{sec:tradeoffs} and \ref{sec:linguistic-content}) computed for different layers of the MLM module of \wvbert (on LibriSpeech dev-clean, scaled embeddings), as illustrated in Figure~\ref{fig:abx-swuggy-sblimp}. In addition, we performed a small subjective evaluation test by listening to a few continuations produced by the different choices. We identified the 7th layer in the MLM module of \wvbert XL and $\codebooksizesemantic=1024$ clusters as our best candidate. 

\looseness -1
\textbf{Acoustic tokens.} We train a SoundStream codec with 12 residual vector quantizer layers with a codebook size of 1024 per layer and 4 convolutional blocks having strides (2, 4, 5, 8). This results in  embeddings sampled at 50 Hz for 16 kHz inputs and a bitrate of 6000\,bps (with 600 tokens per second). As shown in  Table~\ref{tab:token_benchmark}, this model achieves very good reconstruction quality as measured by ViSQOL given the low bitrate. To split the 12 levels of quantization between coarse and fine, we set $Q'=4$ such that we predict the flattened tokens corresponding to the coarse 4 layers in the second stage, whereas the third stage models the fine 8 layers. Hence, the third stage increases the audio bitrate from 2000\,bps to 6000\,bps, which, as shown in Table \ref{tab:token_benchmark}, improves the audio quality significantly. 

\looseness=-1
\textbf{Model.} We use identical decoder-only Transformers in all stages, with 12 layers, 16 attention heads, embedding dimension of 1024, feed-forward layer dimension of 4096 and dropout of 0.1, together with T5-style relative positional embeddings~\cite{t5}, resulting in a model parameter size of 0.3B per stage. During training, we use random cropping to equivalent input lengths of 30, 10 and 3 seconds for the three stages. Furthermore, in the first two stages, we follow the previously proposed practice of removing consecutive repetitions of the semantic tokens~\cite{lakhotia2021generative}. We train each stage on 16 TPUv4s with batch size of 256 for 1M steps.

\textbf{Inference.} We use temperature sampling in all stages, with temperatures of 0.6, 0.8 and 0.6 for the three stages, respectively. We found that these temperature values provide a good trade-off between diversity and semantic consistency of the generated speech. For speech continuation, we use prompts of 3 seconds. For generating the prompts, we truncate samples to the desired prompt length, extract the corresponding \wvbert and SoundStream tokens and use them as conditioning as described in Section~\ref{sec:model-inference}.

\begin{table}[t]
    \centering
    \caption{Character (CER) and word (WER) error rates of the ASR system on audio generated by \ours{} from ground-truth semantic tokens, or by GSLM~\cite{lakhotia2021generative} on ground-truth discrete units. We also report the error rates on the ground-truth audio and its SoundStream reconstruction for reference.}
    \begin{tabular}{cccccc}
    \toprule
     & Original &\begin{tabular}[c]{@{}c@{}}Reconstruction\\with SoundStream\end{tabular}  & \ours{} &\begin{tabular}[c]{@{}c@{}}GSLM~\cite{lakhotia2021generative}\\ unit-to-speech\end{tabular} \\
    \midrule
    CER & $0.8$ & 0.9 & $3.4$ & $2.9$ \\
    WER & $2.5$ & 2.6 & $6.0$ & $6.6$ \\
    \bottomrule
    \end{tabular}
    \label{tab:wer_acoustic_generation}
\end{table}

\begin{table}[t]
    \centering
    \caption{Speaker classification accuracy (\%) on audio generated from ground-truth semantic tokens (``Acoustic generation with \ours{}'') and on continuations of a prompt (``Continuation with \ours{}'').}
    \begin{tabular}{@{}ccc@{}}
    \toprule
 \begin{tabular}[c]{@{}c@{}}Reconstruction\\ with SoundStream\end{tabular} & \begin{tabular}[c]{@{}c@{}}Acoustic generation\\with \ours{} \end{tabular} & \multicolumn{1}{c}{\begin{tabular}[c]{@{}c@{}}Continuation\\ with \ours{} \end{tabular}} \\ \midrule
$100.0$ & $3.2$ & $92.6 $ \\ \bottomrule
    \end{tabular}
    \label{tab:speaker-classification}
\end{table}

\subsection{Information represented by the semantic tokens} \label{sec:exp-semantic}

We investigate the information represented by the different types of tokens to motivate the proposed hierarchical approach based on the separation of semantic and acoustic tokens. In particular, we design experiments for testing the hypothesis that, when modeling speech, the linguistic content is mostly captured by the semantic tokens, while speaker identity and recording conditions are captured by the acoustic tokens.

In the first experiment, we sample the acoustic tokens based on ground-truth semantic tokens extracted from the original speech samples, which corresponds to the ``Acoustic generation'' setup described in Section~\ref{sec:model-inference}. This is done by running the second and third stages, then comparing the linguistic content of the sampled speech to the original speech. Under the hypothesis that the linguistic content is mostly captured by the semantic tokens, the lexical semantics of the sampled and the original speech should coincide. To test this, we perform ASR using a Conformer Transducer-L~\cite{conformertransducer2020} on the generated audio, and calculate the word error rate (WER) and the character error rate (CER) with respect to the original transcripts provided with the data as reference. We use samples from LibriSpeech test-clean with length between 4 and 10 seconds (thus retaining 2.2 hours out of 5.4) and repeat the acoustic generation three times for each sample. For comparison, we also evaluate WER and CER on the same set of samples using the unit-to-speech synthesis module of GSLM~\cite{lakhotia2021generative} as provided by~\texttt{textless-lib}~\cite{textlessLib}. For GSLM resynthesis, we use a vocabulary of 200 tokens derived from HuBERT representations~\cite{hsu2021} which was found to provide the lowest resynthesis error~\cite{lakhotia2021generative}.

\looseness -1
Table~\ref{tab:wer_acoustic_generation} shows the results, where the low WER and CER achieved by \ours{} provide two important insights. First, we can conclude that the semantic content is fully captured by the semantic tokens, as the transcripts obtained from the output of acoustic generation closely follow the original transcripts. Second, the acoustic generation based on sampling SoundStream tokens and decoding them to audio samples preserves good transcription quality. Inspecting the generated samples, we observe that the primary source of errors is the synthesis of proper nouns. A secondary source of errors is the end-of-sentence tokens not being generated at the proper position. Furthermore, since the acoustic generation can synthesize different recording environments, the resulting samples might contain background noise, which also degrades the performance of ASR. Table~\ref{tab:wer_acoustic_generation} also shows that \ours{} performs similarly to GSLM. However, GSLM is trained to synthesize only a single voice in a clean recording environment. Finally, we observe that the error rates of the SoundStream reconstruction are comparable to those of the original audio, suggesting that most of the errors are coming from the mapping of semantic to acoustic tokens. Samples of the acoustic generation are available in the accompanying material.\textsuperscript{\ref{webpage}}

\subsection{Information represented by the acoustic tokens} \label{sec:exp-acoustic}

In the second experiment, we verify the hypothesis that speaker identity and recording conditions are captured by the acoustic tokens. Qualitatively, one can listen to the samples produced by the previous experiment and observe that repeating the sampling of acoustic tokens conditioned on the same semantic tokens results in a wide variety of speakers and recording conditions. 

\looseness -1
In order to perform a quantitative assessment of this observation, we design the following experiment. We train a convolutional network for speaker classification inspired by~\cite{tagliasacchi_self_supervised}, which operates on the log-mel spectrogram of the inputs (25\,ms window length, 10\,ms hop length, 64 mel bins), cropped to 1 second. The network is composed of six convolution blocks, using convolutions along the time and the frequency axes with $3\times1$ and $1\times3$ kernels, followed by ReLU and batch normalization. The number of channels used by each block increases with depth and is equal to [64, 128, 256, 256, 512, 512]. Whenever the number of channels is increased, max pooling with a stride of 2 is also applied along both time and frequency axes. To perform speaker classification on a sequence longer than 1 second, at inference time, we run the classifier on overlapping windows of 1 second with 250\,ms hop length and aggregate the predictions. We train this model on the union of LibriSpeech train-clean-100 and test-clean using the original uncompressed samples, resulting in 291 speakers in total. Then, we randomly split the dataset, with 90\% used for training and 10\% for evaluation. The classifier achieves almost perfect accuracy on the evaluation split of the dataset, and Table~\ref{tab:speaker-classification} shows that it is also robust to lossy compression introduced by SoundStream. 

\looseness -1
To verify that acoustic generation synthesizes different speakers given the same semantic tokens, we run the speaker classifier on the samples generated by \ours{} in that setting. Table~\ref{tab:speaker-classification} shows that, while higher than chance (3.2\% compared to 100 / 291 = 0.3\%), the speaker classification accuracy remains low in this case. We can conclude that the semantic tokens carry little information about the speaker identity, which is instead mostly determined by the acoustic tokens. Furthermore, based on a subjective assessment done by comparing the synthesized samples generated from the same semantic tokens, we observe that rhythm and intonation have only slight variations across different samples, suggesting that prosodic features are captured mostly by the semantic tokens, with some contribution from the acoustic tokens. In addition, we notice a large diversity in the sampled recording conditions, an indication that this characteristic is mainly represented by the acoustic tokens.

\subsection{Probing the linguistic knowledge of \ours{}}%Evaluating linguistic content} 
\label{sec:linguistic-content}
\looseness -1
The previous section shows the linguistic content is captured mostly by modeling of the semantic tokens in the first stage. We now conduct a series of probing experiments to assess the degree of lexical and syntactic knowledge acquired by language modeling of the semantic tokens.
We use two zero-shot metrics, sWUGGY and sBLIMP, introduced in the ZeroResource Challenge 2021~\cite{zerospeech21}. The sWUGGY metric measures whether in a pair of a similar-sounding word and a non-word (e.g., ``brick'' and ``blick''), the model gives a higher probability to the word. In turn, sBLIMP measures how often, according to the model, a grammatically correct sentence has a higher probability than a similar incorrect one (e.g., ``the dogs sleep'' vs.\ ``the dog sleep''). 

\looseness -1
We use the development datasets provided by the organizers of the ZeroResource Challenge 2021, containing 10,000 and 6,300 pairs for the sWUGGY and sBLIMP metrics, respectively, each synthesized using four voices. Following the leaderboard of the challenge, we separately consider the case where the sWUGGY pairs are pre-filtered to contain words that occur in the LibriSpeech data (referred to as the ``in-vocab'' subset).
For both metrics, we identify the positive sample in a pair as the sequence with the higher log-likelihood according to the model.
However, positive examples in the sBLIMP data are on average shorter than their negative counterparts, which can implicitly bias scores towards higher success rates. Thus, we normalize the log-likelihood returned by the model by the sequence length in all experiments.

\looseness -1
We compare the performance of \ours{} to the results reported in the ZeroResource Challenge 2021 leaderboard\footnote{\url{https://zerospeech.com/2021/results.html}} and in the literature. First, we include two text-based toplines which correspond to a BERT model trained on ground-truth phonetic transcriptions, with and without forced alignment, along with a baseline BERT model trained on CPC-derived tokens~\cite{cpc}. Next, we include the leaderboard entry ``Tu Anh et al.'' that corresponds to HuBERT-only model, without an additional language modeling component, which attained the highest sWUGGY and sBLIMP scores in the leaderboard (reported in Nguyen et al.~\cite{nguyentextless2022}). Apart from the baseline, the second-best entry in terms of sWUGGY and sBLIMP is that of Harwath et al.~\cite{Peng2022} which is based on a RoBERTA~\cite{roberta} model trained on top of speech representations obtained with visual grounding. We also add an improved CPC-BERT model by Nguyen et al.~\cite{nguyentextless2022}.

\looseness -1
Unlike \ours{}, the aforementioned models are not causal, so they are not well suited for speech generation. Hence, we also consider causal baselines.  Firstly, we include a variant of GSLM that achieves the best sWUGGY and sBLIMP scores reported in Lakhotia et al.~\cite{lakhotia2021generative}. 
This model is a decoder-only Transformer language model trained using quantized HuBERT representations~\cite{hubert}. Next, we include the entry of van Niekerk et al.~\cite{Niekerk2021}, obtaining the highest scores in the challenge among causal models with an LSTM model trained on CPC-based speech tokens.

\looseness -1
We report the results in Table~\ref{tab:swuggy_sblimp}. Compared to other systems without text supervision, \ours{} achieves the highest sWUGGY scores across both splits. Similarly, it also attains the highest score in the sBLIMP metric, improving by 8\% relative over the previous state-of-the-art (CPC-BERT~\cite{nguyentextless2022}). \ours{} even outperforms a supervised topline using forced aligned phonetic transcriptions.\footnote{Without the log-likelihood normalization discussed above, \ours{} achieves a sBLIMP score of 67.5, outperforming the phone topline.} Overall, our method demonstrates a high ability to model linguistic content without any textual supervision. In particular, it significantly improves over previous work in terms of lexical and syntactic judgement quality.

\begin{table}[t]
    \centering
    \caption{Success rate (\%) on the development sets of sWUGGY and sBLIMP. In bold are best scores among models that do not have text supervision.}
    \begin{tabular}{ccccc}
    \toprule
     Model & \multicolumn{2}{c}{sWUGGY ($\uparrow$)} & sBLIMP ($\uparrow$)   \\
     & all & in-vocab & \\

    \midrule
    \multicolumn{4}{c}{\textit{Text-based toplines}} \\
    \midrule[0.3pt]
    Forced alignment topline~\cite{zerospeech21} & 92.2 & - & 63.7 \\
    Phone topline~\cite{zerospeech21} & 97.9 & - & 66.8 \\
    \midrule
    \multicolumn{4}{c}{\textit{Non-causal}} \\
    \midrule[0.3pt]
    BERT baseline~\cite{zerospeech21} &  67.7 & 75.6 &  56.1 \\
    HuBERT-only~\cite{nguyentextless2022} &  $70.9$ & 79.8 & $59.5$ \\
    Harwath et al.~\cite{Peng2022} & 67.6 & 75.4 &  56.7 \\
    CPC-BERT~\cite{nguyentextless2022} &  - & $80.0$ & $59.9$ \\
    \midrule
    \multicolumn{4}{c}{\textit{Causal}} \\
    \midrule[0.3pt]
    van Niekerk et al.~\cite{Niekerk2021} & 64.3 & 72.3 & 54.0 \\
    GSLM~\cite{lakhotia2021generative} & - &  $68.7$ &  $57.1$\\
    \ours{} & \textbf{71.5} & \textbf{83.7} &  \textbf{64.7}	\\
    \bottomrule
    \end{tabular}
    \label{tab:swuggy_sblimp}
\end{table}

\subsection{Generating coherent continuations} \label{sec:prompt-continuation}
The experiments in Section \ref{sec:linguistic-content} show the capacity of \ours{} to model semantically and syntactically correct linguistic content. For generating convincing continuations, however, we also need coherent acoustic generation. One hallmark of our framework is its ability to continue short prompts of only 3 seconds coherently. 

\looseness -1
To validate the acoustic consistency at the level of speaker identity, we reuse our speaker classifier from Section~\ref{sec:exp-acoustic} and check whether the same speaker is detected in the prompt as in the generated continuation. Concretely, we generate three continuations of 7 seconds for each 3-second prompt, where the prompts are obtained by cropping samples from Librispeech test-clean, whose length is between 4 and 10 seconds. Then, we run the speaker classifier on the sampled continuations (excluding the prompts). The last column of Table~\ref{tab:speaker-classification} shows that the speaker classification accuracy is higher than 92\%, demonstrating that \ours{} generates continuations that strongly preserve the speaker identity. Thus, while semantic tokens carry very little speaker information, prompting \ours{} with both semantic and acoustic tokens allows preserving the speaker identity. 

\subsection{Subjective evaluation}\label{sec:subjective-eval}

\looseness -1
We further validate the result from the previous section by means of a subjective evaluation based on the following task. Raters are asked to listen to a sample of exactly 10 seconds and decide whether it is an original recording of human speech or a synthetic continuation generated by our framework. We use 100 samples in total selected from LibriSpeech test-clean, chosen at random from those with a length of at least 10 seconds, so that we can truncate the length to be exactly 10 seconds without introducing any padding. Half of these samples are ground-truth 10-second utterances, that we compress with SoundStream to match the bitrate of \ours{} outputs, such that compression artifacts cannot be used as cues to detect synthetic audio. From the remaining half, we extract prompts of 3 seconds from the beginning of the samples and generate the corresponding continuations of exactly 7 seconds (resulting in samples of 10 seconds after concatenating with the prompts). We rely on 10 raters screened for proficiency in English and instruct them that the first 3 seconds in each sample is original human speech, and thus their decision should be based on the segment following the first 3 seconds.

\looseness -1
This subjective evaluation task tests at the same time multiple desirable properties: i) the semantic and syntactic correctness of the generated linguistic content; ii) the acoustic coherence of the continuation in context of the prompt (speaker identity, prosody, recording conditions) and iii) the absence of generation artifacts. Based on the 1000 ratings collected, we find that the rate of success for assigning the correct label (original vs. synthesized) is 51.2\%, which, according to a binomial test, is not statistically significantly different ($p=0.23$) from assigning labels uniformly at random (50\% success rate). Since human raters struggle to differentiate short speech samples synthesized by \ours{} from real speech samples in an unpaired setup, the responsible model development practices call for addressing this aspect systematically, which we pursue in the following section. 

\subsection{Detecting synthesized speech} \label{sec:detecting-synthesis}
\looseness -1
We acknowledge that speech generation capabilities of \ours{} carry potential risks, which we further elaborate in Section~\ref{sec:broader-impact}. 
As a way of mitigating such risks, we accompany our framework with a method for detecting whether a speech sequence was synthesized by \ours{}.
To this end, we train a convolutional network with the same architecture as the one described in Section~\ref{sec:exp-acoustic}, but for the binary classification task of differentiating between original samples and continuations generated by \ours{} (excluding the prompt). More precisely, we compare continuations to original samples compressed through SoundStream rather than uncompressed audio, since otherwise i) the task is trivial (the model quickly converges to 100\% accuracy) and ii) eventual compression artifacts would become a confounding factor that would prevent evaluating the generative abilities of \ours{}. For training, we extract the original samples and prompts from LibriSpeech train-clean-100. We train on crops of 1 seconds, and compute predictions on longer sequences with the same approach as described in Section~\ref{sec:exp-acoustic}. On a balanced evaluation set, this model achieves 98.6\% accuracy. This shows that despite being (almost) indistinguishable to human ears as shown by the subjective evaluation presented in Section~\ref{sec:subjective-eval}, continuations generated by \ours{} are very easy to detect with a simple audio classifier.

\subsection{Piano continuation} \label{sec:piano-continuation}
\looseness -1
We demonstrate how our approach extends beyond speech by  generating coherent piano music continuations. For this, we retrain all components of \ours{} on an internal dataset of 40k hours of piano music that includes players from beginner to expert level, and exhibits a wide range of different acoustic conditions, with content ranging from piano scale exercises to famous pieces. The model hyperparameters are identical to the speech continuation setup, except for the acoustic generation stage: we found that a codec with 3 layers of quantization and a larger codebook size of $2^{14}$ per layer already provides high reconstruction quality, so the experiments on piano continuation ignore the third stage and directly predict the 3 levels of acoustic tokens in the second stage.
At inference, we extract a 4-second prompt from the Maestro dataset~\cite{maestro}. The accompanying material\textsuperscript{\ref{webpage}} shows side-by-side comparisons of generations based on acoustic tokens only, or using the full \ours{} framework. While both are of equally high audio quality, analogously to the speech continuation experiments, only the latter display consistent melody and temporal structure. To substantiate this observation, we conduct a subjective evaluation test with 10 raters, in which the raters are asked to express their preference between 15 pairs of continuations (20 seconds each) generated using a model trained on only acoustic tokens and \ours{}, using the same prompt for each pair. The raters preferred the samples produced by \ours{} in 83.3\% of the pairs. This shows that the hierarchical modeling of AudioLM, from semantic to acoustic tokens, not only benefits speech generation by separating linguistic content from speaker identity, but more generally improves audio generation by explicitly disentangling the long-term structure and local acoustic details.

\section{Conclusion}
\looseness -1
We introduce \ours{}, a framework for audio generation that provides both long-term coherence and high audio quality. Relying on a hybrid tokenization scheme of semantic and acoustic tokens, \ours{} performs autoregressive prediction by cascading three stages of language modeling that hierarchically generate audio from the coarsest semantic level up to the finest acoustic details. Experiments on speech generation show that not only~\ours{} can generate syntactically and semantically coherent speech without any text, but also that continuations produced by our model are almost indistinguishable from real speech by humans. We alleviate the risks associated to these realistic continuations by training a classifier which recognizes speech generated by our method with very high accuracy. Furthermore, we show that~\ours{} can generate high-quality piano continuations, demonstrating the benefits of our framework for audio generation beyond speech. This encourages the future extensions to other types of audio (e.g., multilingual speech, polyphonic music, and audio events) as well as integrating~\ours{} into an encoder-decoder framework for conditioned tasks such as text-to-speech or speech-to-speech translation.

\section{Broader Impact} \label{sec:broader-impact}
The ability of \ours{} to synthesize high-quality audio with long-term coherent structure unlocks use-cases ranging from helping people with speech impediments to assisting in composing music. However, there are several risks associated with our model. When modeling speech, \ours{} inherits all concerns about language models for text, such as reflecting the societal biases in the underlying data --- we refer to Chowdhery et al.~\cite{palm} for a detailed discussion on the ethical considerations for text-based language models. Furthermore, the generated speech continuations might not be consistent with the prompt in terms of accent and dialect for underrepresented groups in the training data. 
The ability to continue short speech segments while maintaining speaker identity and prosody can potentially lead to malicious use-cases such as spoofing biometric identification~\cite{asvspoof21} or impersonating a specific speaker~\cite{yourtts}. Therefore, following the responsible AI practices, it is of paramount importance to design mechanisms that safeguard against the misuse of \ours{}. As an important step towards this direction, in Section~\ref{sec:detecting-synthesis} we provide a model for accurately detecting audio synthesized by \ours{}. 

\section{Acknowledgements}
The authors thank John Hershey and Johnny Soraker for their feedback on this work.

\bibliography{main}

\begin{thebibliography}{10}
\providecommand{\url}[1]{#1}
\csname url@samestyle\endcsname
\providecommand{\newblock}{\relax}
\providecommand{\bibinfo}[2]{#2}
\providecommand{\BIBentrySTDinterwordspacing}{\spaceskip=0pt\relax}
\providecommand{\BIBentryALTinterwordstretchfactor}{4}
\providecommand{\BIBentryALTinterwordspacing}{\spaceskip=\fontdimen2\font plus
\BIBentryALTinterwordstretchfactor\fontdimen3\font minus
  \fontdimen4\font\relax}
\providecommand{\BIBforeignlanguage}[2]{{%
\expandafter\ifx\csname l@#1\endcsname\relax
\typeout{** WARNING: IEEEtran.bst: No hyphenation pattern has been}%
\typeout{** loaded for the language `#1'. Using the pattern for}%
\typeout{** the default language instead.}%
\else
\language=\csname l@#1\endcsname
\fi
#2}}
\providecommand{\BIBdecl}{\relax}
\BIBdecl

\bibitem{oord2016wavenet}
A.~van~den Oord, S.~Dieleman, H.~Zen, K.~Simonyan, O.~Vinyals, A.~Graves,
  N.~Kalchbrenner, A.~Senior, and K.~Kavukcuoglu, ``Wave{N}et: A generative
  model for raw audio,'' in \emph{arXiv:1609.03499}, 2016.

\bibitem{kalchbrenner2018wavernn}
N.~Kalchbrenner, E.~Elsen, K.~Simonyan, S.~Noury, N.~Casagrande, E.~Lockhart,
  F.~Stimberg, A.~van~den Oord, S.~Dieleman, and K.~Kavukcuoglu, ``Efficient
  neural audio synthesis,'' in \emph{International Conference on Machine
  Learning (ICML)}, vol.~80.\hskip 1em plus 0.5em minus 0.4em\relax {PMLR},
  2018, pp. 2415--2424.

\bibitem{kumar2019melgan}
K.~Kumar, R.~Kumar, T.~de~Boissiere, L.~Gestin, W.~Z. Teoh, J.~Sotelo,
  A.~de~Brebisson, Y.~Bengio, and A.~Courville, ``{MelGAN}: Generative
  adversarial networks for conditional waveform synthesis,'' in \emph{Advances
  in Neural Information Processing Systems (NeurIPS)}, 2019.

\bibitem{kong2020hifigan}
J.~Kong, J.~Kim, and J.~Bae, ``Hifi-gan: Generative adversarial networks for
  efficient and high fidelity speech synthesis,'' in \emph{Advances in Neural
  Information Processing Systems (NeurIPS)}, 2020.

\bibitem{tagliasacchi2020seanet}
M.~Tagliasacchi, Y.~Li, K.~Misiunas, and D.~Roblek, ``{SEANet}: A multi-modal
  speech enhancement network,'' in \emph{Interspeech}, 2020.

\bibitem{diffwave}
Z.~Kong, W.~Ping, J.~Huang, K.~Zhao, and B.~Catanzaro, ``Diff{W}ave: {A}
  versatile diffusion model for audio synthesis,'' in \emph{International
  Conference on Learning Representations (ICLR)}, 2021.

\bibitem{wavegrad}
N.~Chen, Y.~Zhang, H.~Zen, R.~J. Weiss, M.~Norouzi, and W.~Chan, ``Wave{G}rad:
  Estimating gradients for waveform generation,'' in \emph{International
  Conference on Learning Representations (ICLR)}, 2021.

\bibitem{gpt3}
T.~B. Brown, B.~Mann, N.~Ryder, M.~Subbiah, J.~Kaplan, P.~Dhariwal,
  A.~Neelakantan, P.~Shyam, G.~Sastry, A.~Askell, S.~Agarwal,
  A.~Herbert{-}Voss, G.~Krueger, T.~Henighan, R.~Child, A.~Ramesh, D.~M.
  Ziegler, J.~Wu, C.~Winter, C.~Hesse, M.~Chen, E.~Sigler, M.~Litwin, S.~Gray,
  B.~Chess, J.~Clark, C.~Berner, S.~McCandlish, A.~Radford, I.~Sutskever, and
  D.~Amodei, ``Language models are few-shot learners,'' in \emph{Advances in
  Neural Information Processing Systems (NeurIPS)}, 2020.

\bibitem{gopher}
J.~W. Rae, S.~Borgeaud, T.~Cai, K.~Millican, J.~Hoffmann, F.~Song,
  J.~Aslanides, S.~Henderson, R.~Ring, S.~Young \emph{et~al.}, ``Scaling
  language models: Methods, analysis \& insights from training {G}opher,''
  \emph{arXiv:2112.11446}, 2021.

\bibitem{palm}
A.~Chowdhery, S.~Narang, J.~Devlin, M.~Bosma, G.~Mishra, A.~Roberts, P.~Barham,
  H.~W. Chung, C.~Sutton, S.~Gehrmann, P.~Schuh, K.~Shi, S.~Tsvyashchenko,
  J.~Maynez, A.~B. Rao, P.~Barnes, Y.~Tay, N.~M. Shazeer, V.~Prabhakaran,
  E.~Reif, N.~Du, B.~C. Hutchinson, R.~Pope, J.~Bradbury, J.~Austin, M.~Isard,
  G.~Gur-Ari, P.~Yin, T.~Duke, A.~Levskaya, S.~Ghemawat, S.~Dev,
  H.~Michalewski, X.~Garc{\'i}a, V.~Misra, K.~Robinson, L.~Fedus, D.~Zhou,
  D.~Ippolito, D.~Luan, H.~Lim, B.~Zoph, A.~Spiridonov, R.~Sepassi, D.~Dohan,
  S.~Agrawal, M.~Omernick, A.~M. Dai, T.~S. Pillai, M.~Pellat, A.~Lewkowycz,
  E.~O. Moreira, R.~Child, O.~Polozov, K.~Lee, Z.~Zhou, X.~Wang, B.~Saeta,
  M.~D{\'i}az, O.~Firat, M.~Catasta, J.~Wei, K.~S. Meier-Hellstern, D.~Eck,
  J.~Dean, S.~Petrov, and N.~Fiedel, ``{P}a{LM}: Scaling language modeling with
  {P}athways,'' \emph{arXiv:2204.02311}, 2022.

\bibitem{vit-vqgan2021}
J.~Yu, X.~Li, J.~Y. Koh, H.~Zhang, R.~Pang, J.~Qin, A.~Ku, Y.~Xu, J.~Baldridge,
  and Y.~Wu, ``Vector-quantized image modeling with improved {VQGAN},'' in
  \emph{International Conference on Learning Representations (ICLR)}, 2022.

\bibitem{parti}
J.~Yu, Y.~Xu, J.~Y. Koh, T.~Luong, G.~Baid, Z.~Wang, V.~Vasudevan, A.~Ku,
  Y.~Yang, B.~K. Ayan, B.~Hutchinson, W.~Han, Z.~Parekh, X.~Li, H.~Zhang,
  J.~Baldridge, and Y.~Wu, ``Scaling autoregressive models for content-rich
  text-to-image generation,'' \emph{Transactions on Machine Learning Research
  (TMLR)}, 2022.

\bibitem{zerospeech21}
E.~Dunbar, M.~Bernard, N.~Hamilakis, T.~A. Nguyen, M.~de~Seyssel,
  P.~Roz{\'{e}}, M.~Rivi{\`{e}}re, E.~Kharitonov, and E.~Dupoux, ``The zero
  resource speech challenge 2021: Spoken language modelling,'' in
  \emph{Interspeech}.\hskip 1em plus 0.5em minus 0.4em\relax {ISCA}, 2021, pp.
  1574--1578.

\bibitem{lakhotia2021generative}
K.~Lakhotia, E.~Kharitonov, W.-N. Hsu, Y.~Adi, A.~Polyak, B.~Bolte, T.-A.
  Nguyen, J.~Copet, A.~Baevski, A.~Mohamed \emph{et~al.}, ``On generative
  spoken language modeling from raw audio,'' \emph{Transactions of the
  Association for Computational Linguistics}, vol.~9, pp. 1336--1354, 2021.

\bibitem{attentionvaswani}
A.~Vaswani, N.~Shazeer, N.~Parmar, J.~Uszkoreit, L.~Jones, A.~N. Gomez,
  L.~Kaiser, and I.~Polosukhin, ``Attention is all you need,'' in
  \emph{Advances in Neural Information Processing Systems (NeurIPS)}, 2017, pp.
  5998--6008.

\bibitem{soundstream}
N.~Zeghidour, A.~Luebs, A.~Omran, J.~Skoglund, and M.~Tagliasacchi,
  ``Soundstream: An end-to-end neural audio codec,'' \emph{{IEEE} {ACM} Trans.
  Audio Speech Lang. Process.}, vol.~30, pp. 495--507, 2022.

\bibitem{w2vbert}
Y.~Chung, Y.~Zhang, W.~Han, C.~Chiu, J.~Qin, R.~Pang, and Y.~Wu, ``w2v-bert:
  Combining contrastive learning and masked language modeling for
  self-supervised speech pre-training,'' in \emph{{IEEE} Automatic Speech
  Recognition and Understanding Workshop, {ASRU}}.\hskip 1em plus 0.5em minus
  0.4em\relax {IEEE}, 2021, pp. 244--250.

\bibitem{roberts2022t5x}
A.~Roberts, H.~W. Chung, A.~Levskaya, G.~Mishra, J.~Bradbury, D.~Andor,
  S.~Narang, B.~Lester, C.~Gaffney, A.~Mohiuddin, C.~Hawthorne, A.~Lewkowycz,
  A.~Salcianu, M.~van Zee, J.~Austin, S.~Goodman, L.~B. Soares, H.~Hu,
  S.~Tsvyashchenko, A.~Chowdhery, J.~Bastings, J.~Bulian, X.~Garcia, J.~Ni,
  A.~Chen, K.~Kenealy, J.~H. Clark, S.~Lee, D.~Garrette, J.~Lee-Thorp,
  C.~Raffel, N.~Shazeer, M.~Ritter, M.~Bosma, A.~Passos, J.~Maitin-Shepard,
  N.~Fiedel, M.~Omernick, B.~Saeta, R.~Sepassi, A.~Spiridonov, J.~Newlan, and
  A.~Gesmundo, ``Scaling up models and data with $\texttt{t5x}$ and
  $\texttt{seqio}$,'' \emph{arXiv:2203.17189}, 2022.

\bibitem{bert}
J.~Devlin, M.~Chang, K.~Lee, and K.~Toutanova, ``{BERT:} pre-training of deep
  bidirectional transformers for language understanding,'' in \emph{Conference
  of the North American Chapter of the Association for Computational
  Linguistics: Human Language Technologies, {NAACL-HLT})}.\hskip 1em plus 0.5em
  minus 0.4em\relax Association for Computational Linguistics, 2019, pp.
  4171--4186.

\bibitem{parallel_wavenet}
A.~van~den Oord, Y.~Li, I.~Babuschkin, K.~Simonyan, O.~Vinyals, K.~Kavukcuoglu,
  G.~van~den Driessche, E.~Lockhart, L.~C. Cobo, F.~Stimberg, N.~Casagrande,
  D.~Grewe, S.~Noury, S.~Dieleman, E.~Elsen, N.~Kalchbrenner, H.~Zen,
  A.~Graves, H.~King, T.~Walters, D.~Belov, and D.~Hassabis, ``Parallel
  {W}ave{N}et: Fast high-fidelity speech synthesis,'' in \emph{International
  Conference on Machine Learning (ICML)}, ser. Proceedings of Machine Learning
  Research, vol.~80.\hskip 1em plus 0.5em minus 0.4em\relax {PMLR}, 2018, pp.
  3915--3923.

\bibitem{gantts}
M.~Binkowski, J.~Donahue, S.~Dieleman, A.~Clark, E.~Elsen, N.~Casagrande, L.~C.
  Cobo, and K.~Simonyan, ``High fidelity speech synthesis with adversarial
  networks,'' in \emph{International Conference on Learning Representations
  (ICLR)}, 2020.

\bibitem{vqvae}
A.~van~den Oord, O.~Vinyals, and K.~Kavukcuoglu, ``Neural discrete
  representation learning,'' in \emph{Advances in Neural Information Processing
  Systems (NeurIPS)}, 2017, pp. 6306--6315.

\bibitem{vqvae2}
A.~Razavi, A.~van~den Oord, and O.~Vinyals, ``Generating diverse high-fidelity
  images with {VQ-VAE-2},'' in \emph{Advances in Neural Information Processing
  Systems (NeurIPS)}, 2019, pp. 14\,837--14\,847.

\bibitem{softhardquantization}
E.~Agustsson, F.~Mentzer, M.~Tschannen, L.~Cavigelli, R.~Timofte, L.~Benini,
  and L.~V. Gool, ``Soft-to-hard vector quantization for end-to-end learned
  compression of images and neural networks,'' \emph{Advances in Neural
  Information Processing Systems (NeurIPS)}, 2017.

\bibitem{kankanahalli2018speechdnn}
S.~Kankanahalli, ``End-to-end optimized speech coding with deep neural
  networks,'' in \emph{IEEE International Conference on Acoustics, Speech and
  Signal Processing (ICASSP)}, 2018, pp. 2521--2525.

\bibitem{zhen2019cascaded}
K.~Zhen, J.~Sung, M.~S. Lee, S.~Beack, and M.~Kim, ``Cascaded cross-module
  residual learning towards lightweight end-to-end speech coding,'' in
  \emph{Interspeech}.\hskip 1em plus 0.5em minus 0.4em\relax {ISCA}, 2019, pp.
  3396--3400.

\bibitem{petermann2021harp}
D.~Petermann, S.~Beack, and M.~Kim, ``{HARP}-{N}et: Hyper-autoencoded
  reconstruction propagation for scalable neural audio coding,''
  \emph{arXiv:2107.10843}, 2021.

\bibitem{cpc}
A.~van~den Oord, Y.~Li, and O.~Vinyals, ``Representation learning with
  contrastive predictive coding,'' \emph{arXiv:1807.03748}, 2018.

\bibitem{wav2vec2}
A.~Baevski, H.~Zhou, A.~Mohamed, and M.~Auli, ``wav2vec 2.0: {A} framework for
  self-supervised learning of speech representations,''
  \emph{arXiv:2006.11477}, 2020.

\bibitem{cola}
A.~Saeed, D.~Grangier, and N.~Zeghidour, ``Contrastive learning of
  general-purpose audio representations,'' in \emph{{IEEE} International
  Conference on Acoustics, Speech and Signal Processing (ICASSP)}.\hskip 1em
  plus 0.5em minus 0.4em\relax {IEEE}, 2021, pp. 3875--3879.

\bibitem{contrastive_kharitonov}
E.~Kharitonov, M.~Rivi{\`{e}}re, G.~Synnaeve, L.~Wolf, P.~Mazar{\'{e}},
  M.~Douze, and E.~Dupoux, ``Data augmenting contrastive learning of speech
  representations in the time domain,'' in \emph{{IEEE} Spoken Language
  Technology Workshop, {SLT} 2021, Shenzhen, China, January 19-22, 2021}.\hskip
  1em plus 0.5em minus 0.4em\relax {IEEE}, 2021, pp. 215--222.

\bibitem{vqwav2vec}
A.~Baevski, S.~Schneider, and M.~Auli, ``vq-wav2vec: Self-supervised learning
  of discrete speech representations,'' in \emph{International Conference on
  Learning Representations (ICLR)}, 2020.

\bibitem{hubert}
W.~Hsu, B.~Bolte, Y.~H. Tsai, K.~Lakhotia, R.~Salakhutdinov, and A.~Mohamed,
  ``Hubert: Self-supervised speech representation learning by masked prediction
  of hidden units,'' \emph{{IEEE} {ACM} Trans. Audio Speech Lang. Process.},
  vol.~29, pp. 3451--3460, 2021.

\bibitem{pasad2021}
A.~Pasad, J.-C. Chou, and K.~Livescu, ``Layer-wise analysis of a
  self-supervised speech representation model,'' in \emph{2021 IEEE Automatic
  Speech Recognition and Understanding Workshop (ASRU)}, 2021, pp. 914--921.

\bibitem{lamda}
R.~Thoppilan, D.~D. Freitas, J.~Hall, N.~Shazeer, A.~Kulshreshtha, H.~Cheng,
  A.~Jin, T.~Bos, L.~Baker, Y.~Du, Y.~Li, H.~Lee, H.~S. Zheng, A.~Ghafouri,
  M.~Menegali, Y.~Huang, M.~Krikun, D.~Lepikhin, J.~Qin, D.~Chen, Y.~Xu,
  Z.~Chen, A.~Roberts, M.~Bosma, Y.~Zhou, C.~Chang, I.~Krivokon, W.~Rusch,
  M.~Pickett, K.~S. Meier{-}Hellstern, M.~R. Morris, T.~Doshi, R.~D. Santos,
  T.~Duke, J.~Soraker, B.~Zevenbergen, V.~Prabhakaran, M.~Diaz, B.~Hutchinson,
  K.~Olson, A.~Molina, E.~Hoffman{-}John, J.~Lee, L.~Aroyo, R.~Rajakumar,
  A.~Butryna, M.~Lamm, V.~Kuzmina, J.~Fenton, A.~Cohen, R.~Bernstein,
  R.~Kurzweil, B.~Aguera{-}Arcas, C.~Cui, M.~Croak, E.~H. Chi, and Q.~Le,
  ``La{MDA}: Language models for dialog applications,''
  \emph{arXiv:2201.08239}, 2022.

\bibitem{openaicodex}
M.~Chen, J.~Tworek, H.~Jun, Q.~Yuan, H.~P. de~Oliveira~Pinto, J.~Kaplan,
  H.~Edwards, Y.~Burda, N.~Joseph, G.~Brockman, A.~Ray, R.~Puri, G.~Krueger,
  M.~Petrov, H.~Khlaaf, G.~Sastry, P.~Mishkin, B.~Chan, S.~Gray, N.~Ryder,
  M.~Pavlov, A.~Power, L.~Kaiser, M.~Bavarian, C.~Winter, P.~Tillet, F.~P.
  Such, D.~Cummings, M.~Plappert, F.~Chantzis, E.~Barnes, A.~Herbert{-}Voss,
  W.~H. Guss, A.~Nichol, A.~Paino, N.~Tezak, J.~Tang, I.~Babuschkin, S.~Balaji,
  S.~Jain, W.~Saunders, C.~Hesse, A.~N. Carr, J.~Leike, J.~Achiam, V.~Misra,
  E.~Morikawa, A.~Radford, M.~Knight, M.~Brundage, M.~Murati, K.~Mayer,
  P.~Welinder, B.~McGrew, D.~Amodei, S.~McCandlish, I.~Sutskever, and
  W.~Zaremba, ``Evaluating large language models trained on code,''
  \emph{arXiv:2107.03374}, 2021.

\bibitem{symbolic_math_lample}
G.~Lample and F.~Charton, ``Deep learning for symbolic mathematics,'' in
  \emph{International Conference on Learning Representations (ICLR)}, 2020.

\bibitem{t5}
C.~Raffel, N.~Shazeer, A.~Roberts, K.~Lee, S.~Narang, M.~Matena, Y.~Zhou,
  W.~Li, and P.~J. Liu, ``Exploring the limits of transfer learning with a
  unified text-to-text transformer,'' \emph{JMLR}, vol.~21, pp. 140:1--140:67,
  2020.

\bibitem{routing_transformer}
A.~Roy, M.~Saffar, A.~Vaswani, and D.~Grangier, ``Efficient content-based
  sparse attention with routing transformers,'' \emph{Trans. Assoc. Comput.
  Linguistics}, vol.~9, pp. 53--68, 2021.

\bibitem{performer}
K.~M. Choromanski, V.~Likhosherstov, D.~Dohan, X.~Song, A.~Gane,
  T.~Sarl{\'{o}}s, P.~Hawkins, J.~Q. Davis, A.~Mohiuddin, L.~Kaiser, D.~B.
  Belanger, L.~J. Colwell, and A.~Weller, ``Rethinking attention with
  performers,'' in \emph{International Conference on Learning Representations
  (ICLR)}, 2021.

\bibitem{perceiverar2022}
C.~Hawthorne, A.~Jaegle, C.~Cangea, S.~Borgeaud, C.~Nash, M.~Malinowski,
  S.~Dieleman, O.~Vinyals, M.~M. Botvinick, I.~Simon, H.~Sheahan, N.~Zeghidour,
  J.~Alayrac, J.~Carreira, and J.~H. Engel, ``General-purpose, long-context
  autoregressive modeling with {P}erceiver {AR},'' in \emph{International
  Conference on Machine Learning (ICML)}, ser. Proceedings of Machine Learning
  Research, vol. 162.\hskip 1em plus 0.5em minus 0.4em\relax {PMLR}, 2022, pp.
  8535--8558.

\bibitem{tamingtransformers}
P.~Esser, R.~Rombach, and B.~Ommer, ``Taming transformers for high-resolution
  image synthesis,'' in \emph{{IEEE} Conference on Computer Vision and Pattern
  Recognition (CVPR)}.\hskip 1em plus 0.5em minus 0.4em\relax Computer Vision
  Foundation / {IEEE}, 2021, pp. 12\,873--12\,883.

\bibitem{tats}
S.~Ge, T.~Hayes, H.~Yang, X.~Yin, G.~Pang, D.~Jacobs, J.~Huang, and D.~Parikh,
  ``Long video generation with time-agnostic {VQGAN} and time-sensitive
  transformer,'' \emph{arXiv:2204.03638}, 2022.

\bibitem{jukebox}
P.~Dhariwal, H.~Jun, C.~Payne, J.~W. Kim, A.~Radford, and I.~Sutskever,
  ``Jukebox: {A} generative model for music,'' \emph{arXiv:2005.00341}, 2020.

\bibitem{pgslm}
E.~Kharitonov, A.~Lee, A.~Polyak, Y.~Adi, J.~Copet, K.~Lakhotia, T.~A. Nguyen,
  M.~Rivi{\`{e}}re, A.~Mohamed, E.~Dupoux, and W.~Hsu, ``Text-free
  prosody-aware generative spoken language modeling,'' in \emph{Proceedings of
  the 60th Annual Meeting of the Association for Computational Linguistics
  (ACL)}.\hskip 1em plus 0.5em minus 0.4em\relax Association for Computational
  Linguistics, 2022, pp. 8666--8681.

\bibitem{textlessLib}
E.~Kharitonov, J.~Copet, K.~Lakhotia, T.~A. Nguyen, P.~Tomasello, A.~Lee,
  A.~Elkahky, W.~Hsu, A.~Mohamed, E.~Dupoux, and Y.~Adi, ``textless-lib: a
  library for textless spoken language processing,'' \emph{arXiv:2202.07359},
  2022.

\bibitem{conformer}
A.~Gulati, J.~Qin, C.~Chiu, N.~Parmar, Y.~Zhang, J.~Yu, W.~Han, S.~Wang,
  Z.~Zhang, Y.~Wu, and R.~Pang, ``Conformer: Convolution-augmented transformer
  for speech recognition,'' in \emph{Interspeech}.\hskip 1em plus 0.5em minus
  0.4em\relax {ISCA}, 2020, pp. 5036--5040.

\bibitem{xtremes}
A.~Conneau, A.~Bapna, Y.~Zhang, M.~Ma, P.~von Platen, A.~Lozhkov, C.~Cherry,
  Y.~Jia, C.~Rivera, M.~Kale, D.~van Esch, V.~Axelrod, S.~Khanuja, J.~H. Clark,
  O.~Firat, M.~Auli, S.~Ruder, J.~Riesa, and M.~Johnson, ``{XTREME-S:}
  evaluating cross-lingual speech representations,'' \emph{arXiv:2203.10752},
  2022.

\bibitem{hines2015visqol}
A.~Hines, J.~Skoglund, A.~C. Kokaram, and N.~Harte, ``{ViSQOL}: an objective
  speech quality model,'' \emph{EURASIP Journal on Audio, Speech, and Music
  Processing}, vol. 2015, no.~1, pp. 1--18, 2015.

\bibitem{chinen2020visqol}
M.~Chinen, F.~S.~C. Lim, J.~Skoglund, N.~Gureev, F.~O'Gorman, and A.~Hines,
  ``{ViSQOL v3:} an open source production ready objective speech and audio
  metric,'' in \emph{Twelfth International Conference on Quality of Multimedia
  Experience (QoMEX)}, 2020, pp. 1--6.

\bibitem{abx}
T.~Schatz, V.~Peddinti, F.~R. Bach, A.~Jansen, H.~Hermansky, and E.~Dupoux,
  ``Evaluating speech features with the minimal-pair {ABX} task: analysis of
  the classical {MFC/PLP} pipeline,'' in \emph{Interspeech}.\hskip 1em plus
  0.5em minus 0.4em\relax {ISCA}, 2013, pp. 1781--1785.

\bibitem{schatz2016abx}
T.~Schatz, ``Abx-discriminability measures and applications. (mesures de
  discriminabilit{\'{e}} {ABX} et applications),'' Ph.D. dissertation, Pierre
  and Marie Curie University, Paris, France, 2016.

\bibitem{librilight}
J.~Kahn, M.~Rivi{\`{e}}re, W.~Zheng, E.~Kharitonov, Q.~Xu, P.~Mazar{\'{e}},
  J.~Karadayi, V.~Liptchinsky, R.~Collobert, C.~Fuegen, T.~Likhomanenko,
  G.~Synnaeve, A.~Joulin, A.~Mohamed, and E.~Dupoux, ``Libri-light: {A}
  benchmark for {ASR} with limited or no supervision,'' in \emph{{IEEE}
  International Conference on Acoustics, Speech and Signal Processing
  (ICASSP)}.\hskip 1em plus 0.5em minus 0.4em\relax {IEEE}, 2020, pp.
  7669--7673.

\bibitem{librispeech}
V.~Panayotov, G.~Chen, D.~Povey, and S.~Khudanpur, ``Librispeech: An {ASR}
  corpus based on public domain audio books,'' in \emph{{IEEE} International
  Conference on Acoustics, Speech and Signal Processing (ICASSP)}.\hskip 1em
  plus 0.5em minus 0.4em\relax {IEEE}, 2015, pp. 5206--5210.

\bibitem{librilight-6k}
M.~Rivi{\`{e}}re and E.~Dupoux, ``Towards unsupervised learning of speech
  features in the wild,'' in \emph{{IEEE} Spoken Language Technology Workshop
  (SLT)}.\hskip 1em plus 0.5em minus 0.4em\relax {IEEE}, 2021, pp. 156--163.

\bibitem{conformertransducer2020}
A.~Gulati, J.~Qin, C.~Chiu, N.~Parmar, Y.~Zhang, J.~Yu, W.~Han, S.~Wang,
  Z.~Zhang, Y.~Wu, and R.~Pang, ``Conformer: Convolution-augmented transformer
  for speech recognition,'' in \emph{Interspeech}.\hskip 1em plus 0.5em minus
  0.4em\relax {ISCA}, 2020, pp. 5036--5040.

\bibitem{hsu2021}
W.~Hsu, Y.~H. Tsai, B.~Bolte, R.~Salakhutdinov, and A.~Mohamed, ``Hubert: How
  much can a bad teacher benefit {ASR} pre-training?'' in \emph{{IEEE}
  International Conference on Acoustics, Speech and Signal Processing
  (ICASSP)}.\hskip 1em plus 0.5em minus 0.4em\relax {IEEE}, 2021, pp.
  6533--6537.

\bibitem{tagliasacchi_self_supervised}
M.~Tagliasacchi, B.~Gfeller, F.~de~Chaumont~Quitry, and D.~Roblek,
  ``Self-supervised audio representation learning for mobile devices,''
  \emph{arXiv:1905.11796}, 2019.

\bibitem{nguyentextless2022}
T.~A. Nguyen, B.~Sagot, and E.~Dupoux, ``Are discrete units necessary for
  spoken language modeling?'' \emph{arXiv:2203.05936}, 2022.

\bibitem{Peng2022}
P.~Peng and D.~Harwath, ``Self-supervised representation learning for speech
  using visual grounding and masked language modeling,''
  \emph{arXiv:2202.03543}, 2022.

\bibitem{roberta}
Y.~Liu, M.~Ott, N.~Goyal, J.~Du, M.~Joshi, D.~Chen, O.~Levy, M.~Lewis,
  L.~Zettlemoyer, and V.~Stoyanov, ``Roberta: {A} robustly optimized {BERT}
  pretraining approach,'' \emph{arXiv:1907.11692}, 2019.

\bibitem{Niekerk2021}
B.~van Niekerk, L.~Nortje, M.~Baas, and H.~Kamper, ``Analyzing speaker
  information in self-supervised models to improve zero-resource speech
  processing,'' in \emph{Interspeech}.\hskip 1em plus 0.5em minus 0.4em\relax
  {ISCA}, 2021, pp. 1554--1558.

\bibitem{maestro}
C.~Hawthorne, A.~Stasyuk, A.~Roberts, I.~Simon, C.~A. Huang, S.~Dieleman,
  E.~Elsen, J.~H. Engel, and D.~Eck, ``Enabling factorized piano music modeling
  and generation with the {MAESTRO} dataset,'' in \emph{International
  Conference on Learning Representations (ICLR)}, 2019.

\bibitem{asvspoof21}
H.~Delgado, N.~W.~D. Evans, T.~Kinnunen, K.~A. Lee, X.~Liu, A.~Nautsch,
  J.~Patino, M.~Sahidullah, M.~Todisco, X.~Wang, and J.~Yamagishi, ``Asvspoof
  2021: Automatic speaker verification spoofing and countermeasures challenge
  evaluation plan,'' \emph{arXiv:2109.00535}, 2021.

\bibitem{yourtts}
E.~Casanova, J.~Weber, C.~D. Shulby, A.~C. J{\'{u}}nior, E.~G{\"{o}}lge, and
  M.~A. Ponti, ``Yourtts: Towards zero-shot multi-speaker {TTS} and zero-shot
  voice conversion for everyone,'' in \emph{International Conference on Machine
  Learning (ICML)}, ser. Proceedings of Machine Learning Research, vol.
  162.\hskip 1em plus 0.5em minus 0.4em\relax {PMLR}, 2022, pp. 2709--2720.

\end{thebibliography}
\bibliographystyle{IEEEtran}

\end{document}